\documentclass{article}
\usepackage{spconf,amsmath,amsfonts,graphicx}
\usepackage{caption}
\usepackage{subcaption}
\usepackage{dsfont}
\usepackage{xcolor}
\usepackage{soul}
\usepackage{hyperref}

\usepackage{algorithm,algpseudocode}

\DeclareMathOperator*{\argmin}{arg\,min}

\newcommand{\rank}{\mathrm{rank}}
\newcommand{\bR}{\mathbb{R}}

\newcommand{\HTr}[1]{\mathrm{H}^\mathrm{SVD}_r\left(#1\right)}
\newcommand{\HTs}[1]{\mathrm{H}^{\ell_0}_s\left(#1\right)}

\newcommand{\Rev}[1]{#1} 
\newcommand{\RevT}[1]{#1} 

\title{Low-rank plus sparse trajectory decomposition \\for direct exoplanet imaging}
\name{Simon Vary, Hazan Daglayan, Laurent Jacques, and Pierre-Antoine Absil
\thanks{This work was supported by the Fonds de la Recherche Scientifique - FNRS and the Fonds Wetenschappelijk Onderzoek - Vlaanderen under EOS Project no. 30468160. Simon Vary is a beneficiary of the FSR Incoming Post-doctoral Fellowship. Laurent Jacques is supported by Belgian National Science Foundation (F.R.S.-FNRS).}
}
\address{
    ICTEAM/INMA\\ UCLouvain, Belgium
 }
\begin{document}
%
\maketitle
\begin{abstract}
We propose a direct imaging method for the detection of exoplanets based on a combined low-rank plus structured sparse model. For this task, we develop a dictionary of possible \Rev{effective} circular trajectories a planet can take \Rev{during the observation time}, elements of which can be efficiently computed using rotation and convolution operation. We design a simple \Rev{alternating} iterative hard-thresholding algorithm \Rev{that jointly promotes a low-rank background and a sparse exoplanet foreground,} to solve the non-convex optimisation problem. \Rev{The experimental comparison on the $\beta$-Pictoris exoplanet benchmark dataset} shows that our method has the potential to outperform the widely used Annular PCA for specific planet light intensities in terms of the Receiver operating characteristic (ROC) curves.
\end{abstract}
\section{Introduction}
\label{sec:intro}

Identification, confirmation, and characterization of planets in nearby solar systems is a key challenge in astronomy. The number of confirmed exoplanets recently surpassed 5000 planets, most of these accomplished by \emph{indirect imaging} methods in the past 30 years \cite{NASA_catalog}. 

Indirect imaging methods are based on measuring the effect a planet has on the starlight reaching the Earth, such as when a planet obstructs the light of the star, called the \emph{transit method}. As such, they are  limited to specific alignment between the planet, the star, and the observer, and biased towards close and massive planets orbiting old quiet stars \cite{Daglayan2022Likelihood}. 

In comparison, \emph{directly imaging} exoplanets is a much more difficult task. The challenge arises mostly due to their extremely faint light source compared to their parent star, which can be sometimes about a billion times as bright \Rev{\cite{GomezGonzalez2016Lowranka}}. This requires a telescope that can capture images of both high-resolution and high contrast, \Rev{which is generally done from the ground, using the largest optical telescopes equipped with dedicated instrumentation.} \Rev{However, images acquired with ground-based telescopes are subject to noise caused by atmospheric turbulence resulting in a high-intensity noise called \emph{quasi-static speckles}. Inconveniently, the quasi-static speckled noise is very similar in shape and intensity to the planet companions making their detection by direct imaging a challenging task.}

\emph{Angular differential imaging} (ADI) is the leading method for direct imaging of exoplanets, which, when combined with clever image post-processing techniques, can remove most of the quasi-static speckled noise \cite{Marois2006Angular}. Its idea is to take a sequence of images over a single night of observation without compensating for the Earth's rotation. As a result, planet companions in the captured sequence of images follow circular trajectories of known angular velocity, while the star and the quasi-static speckles, which get introduced only when the light passes through the Earth's atmosphere, remain fixed with respect to the telescope. From a datacube of such observations, it then becomes possible to estimate the quasi-static speckles and subtract them from the original observations. 

Many of the most successful methods for removal of the quasi-static speckled noise are based on low-rank matrix models, such as principal component analysis (PCA)~\cite{Amara2012Pynpoint,Soummer2012Detection}, its annular version (AnnPCA) \cite{GomezGonzalez2017Vip}, the low-rank plus sparse method (LLSG)~\cite{GomezGonzalez2016Lowranka}, and the morphological component analysis (MAYO) \cite{Pairet2021MAYONNAISEa}. The rationale behind the low-rank models is that the bright, quasi-static speckles that appear as static elements with slowly changing light intensity, are captured by the first few principal components, while the higher-rank moving planets are excluded from the model.

\begin{figure*}[t!]
    \centering
    \begin{subfigure}{\textwidth}
        \centering
        \includegraphics[width = 0.95\textwidth]{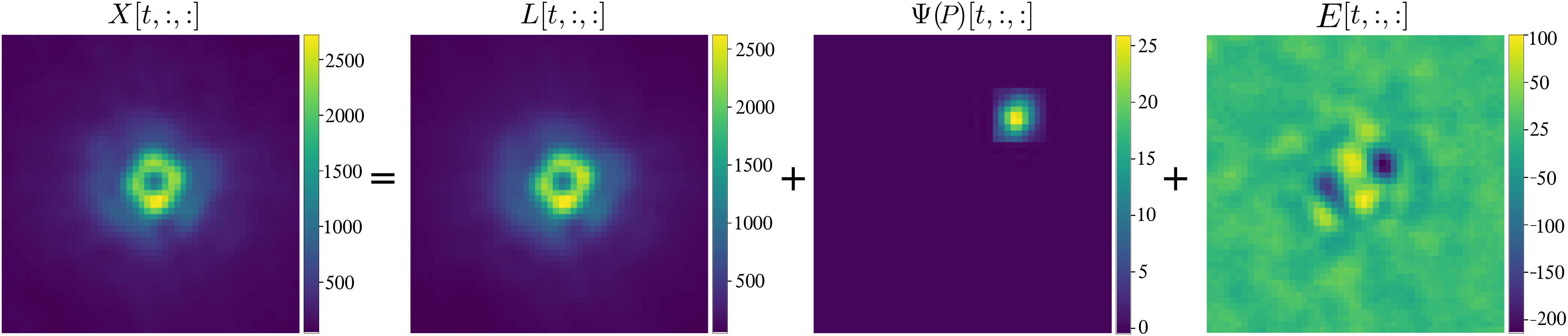}
    \end{subfigure}
    \caption{The result of the low-rank plus \Rev{sparse trajectory} (LRPT) \Rev{method applied to the $\beta$-Pictoris dataset with the real planet}. We show a single frame $t$ of the low-rank component, which captures the quasi-static speckled noise, and the $\Psi(P)$ component which detects the rotating planet. \label{fig:trajectorlet_decomposition}}
\end{figure*}

Following the subtraction of the quasi-static speckled background, the typical next step in the ADI pipeline is the estimation of the light intensity of possible planet companions, referred to as the \emph{flux estimation}. Classical approaches take into account the point spread function (PSF) of \Rev{the optical device} with the known angular velocity and are median-based \cite{Marois2006Angular} or likelihood based \cite{Mugnier2009Optimal, Daglayan2022Likelihood}. More recently, MAYO proposed to employ an image sparsifying transform to model general rotating objects which allows it to detect also circumstellar discs \cite{Pairet2021MAYONNAISEa}.

A key image processing challenge in the ADI pipeline is to be able to distinguish the moving planets from the quasi-static speckles as they are of similar intensities and resolution. A limitation of the ADI pipeline is that it works in steps: it assumes a low-rank model for the quasi-static speckles, subtracts the result, and then it proceeds to estimate the flux \Rev{of possible planets}.

\begin{figure}[b]
    \centering
    \includegraphics[width = .49\textwidth]{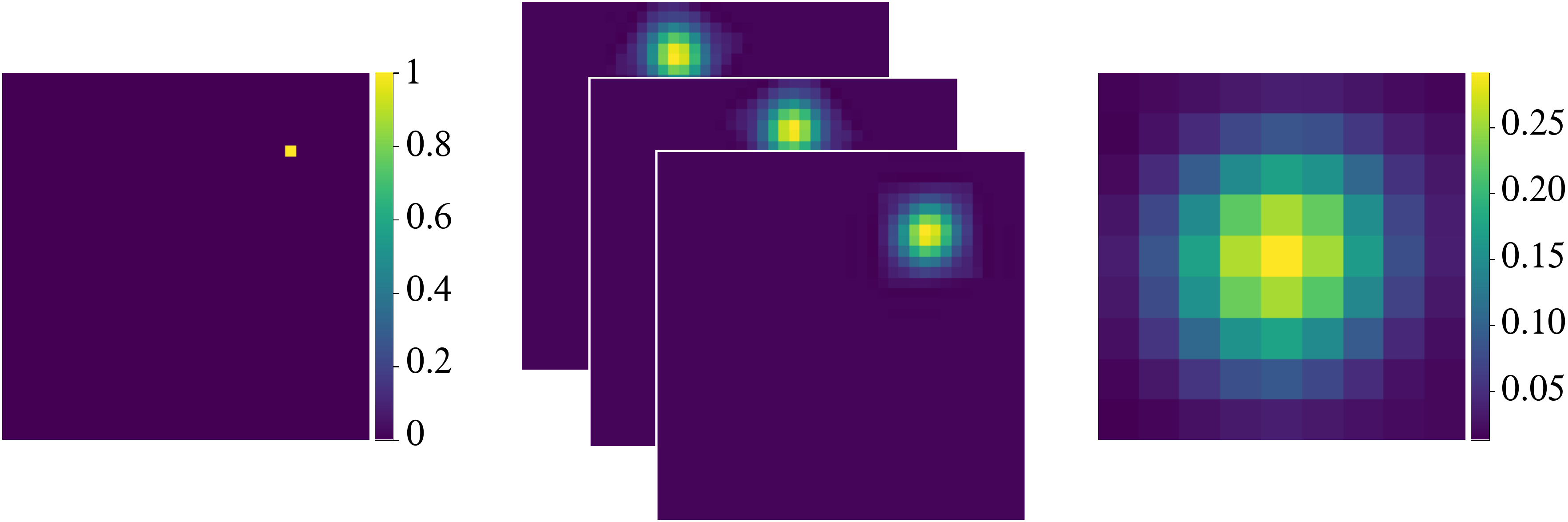}
    \caption{Illustration of the trajectorlet transform: (left) depicts a matrix $P\in\bR^{35\times 35}$ with $P_{25,25} = 1$ and $0$ otherwise, (center) shows $\Psi(P)$, i.e., a single element of the dictionary denoted $\Psi_{25,25}$, (right) shows the PSF of the \Rev{optical device}.\label{fig:trajectorlet}}
\end{figure}

Methods such as LLSG~\cite{GomezGonzalez2016Lowranka} and the MAYO pipeline \cite{Pairet2021MAYONNAISEa} have been developed with these issues in mind. In LLSG, the authors propose to perform low-rank plus sparse decomposition, which prevents accidentally subtracting the planet from the datacube by accounting for sparse outliers that can capture the planet. However, the issue here is that the sparse component could also capture a quasi-static speckle since it does not assume that the planet needs to move smoothly in a circular trajectory. 
The most similar to our approach is the MAYO method \cite{Pairet2021MAYONNAISEa}, which assumes both a low-rank model for the quasi-speckled background \Rev{and} a circular trajectory of the planet. However, whereas MAYO performs these tasks as part of a multi-step pipeline, in this work, we employ an alternating hard thresholding algorithm \cite{Tanner2022Compressed} that solves them at once as part of a single optimization problem.

In this paper, we propose to employ a \emph{low-rank plus sparse trajectory} method \Rev{(LRPT)} that estimates the low-rank component while trying to find a small number of viable planet trajectories based on a low-rank plus sparse decomposition \cite{Tanner2022Compressed}. We develop a linear transform, referred to as the \emph{trajectorlet transform}, which takes into account the knowledge of possible planet trajectories and can be computed efficiently using convolution operations and rotation.

The paper is structured as follows. The following Section~\ref{sec:method} introduces the low-rank plus sparse model and the trajectorlet transform. In Section~\ref{sec:numerics}, we present experimental results and comparisons of the method with other standard methods. Finally, Section~\ref{sec:conclusion} concludes the paper and suggests further research directions using this model.

\section{Direct imaging with low-rank plus sparse model}\label{sec:method}

Let $X\in\bR^{T\times N^2}$ be the matrix of observations whose rows are the $N \times N$ image frames of the video sequence with $T$ frames that contains an unknown small number of planets as well as the quasi-static speckle noise. Our model for $X$, \Rev{depicted in Fig.~\ref{fig:trajectorlet_decomposition},} is expressed as
\begin{equation} \label{eq:model}
\begin{gathered}
    X = L + \Psi(P) + E \\
    \mathrm{s.t.}\qquad \rank{(L)}\leq r,\quad \|P\|_0\leq s,\quad P_{i,j} \geq 0, 
\end{gathered}
\end{equation}
where $L\in\bR^{T\times N^2}$ is the rank-$r$ component which models the quasi-static speckles, $\Psi(P)$ is the component containing the rotating planets with $P\in\bR^{N\times N}$ being a non-negative $s$-sparse matrix and $\Psi:\bR^{N\times N}\rightarrow \bR^{T\times N^2}$ is a linear \emph{trajectorlet} transform specified below. The final component $E$ is the residual noise which we assume to be normally distributed. 

\subsection{Trajectorlet transform}\label{subsec:trajectorlet}
The trajectorlet transform $\Psi:\mathbb{R}^{N\times N} \rightarrow \mathbb{R}^{T \times N^2}$ is a linear mapping that expresses possible trajectories a planet can take in the $N\times N$ frame according to the known angular velocity and the point spread function (PSF) of \Rev{the optical device}. 

The linear transform $\Psi(\cdot)$ is represented by a dictionary of $N^2$ elements in $\bR^{T \times N^2}$ which we denote $\Psi_{i,j}$ for $\forall i,j \in [N] \times [N]$. A single item $\Psi_{i,j}$ of the dictionary is generated by placing the PSF of the \Rev{optical device} at the position $i,j$ on the first frame and rotating it according to the known angular velocity of the telescope on the remaining frames; see Fig.~\ref{fig:trajectorlet} for an illustration. \Rev{We normalize the elements of $\Psi_{i,j}$ to produce a unit norm frame}. \Rev{The matrix-vector product and the adjoint product are defined as
\begin{equation}
    \Psi(P) = \sum_{i,j = 1}^{N, N} \Psi_{i,j}\, P_{i,j} \quad 
    \Psi^*(X) = \sum_{i,j = 1}^{N, N} \left\langle
    \Psi_{i,j},\, X 
    \right\rangle \mathds{1}_{i,j}, \label{eq:trajectorlet}
\end{equation}
where $\langle \cdot, \cdot \rangle$ is the trace inner product and $\mathds{1}_{i,j}\in\bR^{N\times N}$ denotes a canonical basis matrix with a single one at index $i,j$ and zero otherwise.}

\Rev{The underlying structure of the transform allows us to compute the operations in \eqref{eq:trajectorlet} in an computationally efficient way through the use of a rotation and a convolution operation.}  To compute the forward operation $\Psi(P)$, we apply the convolution of $P\in\bR^{N\times N}$ with the PSF of the planet as the kernel and by placing the rotated copy of the result along the $T$ frames. The adjoint operation $\Psi^*(X)$ can be computed by reversing the operations: first, we derotate the rows of $X\in\bR^{T\times N^2}$ summing them into an $N^2$ vector, we unfold it into a $N\times N$ matrix and perform \Rev{the adjoint of the convolution} with the PSF of the planet as the kernel. Detailed implementation using the \texttt{scipy.sparse} API \cite{Virtanen2020scipy} can be found in the code provided with our paper\footnote{Link to the code \url{https://github.com/hazandaglayan/trajectorlets}.\label{footnote:code}}. Note that the rotation and convolution procedure has also been used in MAYO \cite{Pairet2021MAYONNAISEa}.

\subsection{Low-rank plus structured sparse model}\label{subsec:trajectorlet}
We compute the best unbiased estimator for the model in \eqref{eq:model} by solving the following non-convex optimisation problem
\begin{equation}
\begin{gathered}
     \min_{L \in \bR^{T\times N^2},\, P\in\bR^{N\times N}} \Rev{\frac{1}{2}\left\| X - \left(L + \Psi(P) \right)\right\|_F^2}, \\
     \mathrm{s.t.}\qquad \rank{(L)} \leq r,\quad \|P\|_0\leq s,\quad P_{i,j} \geq 0, 
\end{gathered}\label{eq:optimization}
\end{equation}
by applying a modification of the Normalized Alternating Hard Thresholding (NAHT) algorithm \cite{Tanner2022Compressed}, which is shown in Algorithm~\ref{algo:naht}. The method solves \eqref{eq:optimization} by taking alternating projected gradient steps. \Rev{In line 5, NAHT computes an estimate of the quasi-static background by projecting $X-\Psi(P^j)$ on the set of rank-$r$ matrices \RevT{with $\HTr{\cdot}$ that performs the randomized singular value decomposition and keeps only $r$ largest singular values of the input matrix \cite{Halko2011Finding}}. Next, in line 6, it computes a gradient direction of the objective \eqref{eq:optimization} in respect to $P$ based on the current residual map. The algorithm estimates the adaptive step-size in line 7, where $R^j_{\Omega^{j-1}}$ denotes the matrix $R^j$ with entries kept only on the support set $\Omega^{j-1}$ and zeroes otherwise. The step-size corresponds to the optimal gradient step-size when the current iterate $P^j$ has the same support as the optimal sparse solution \cite{Blumensath2010Normalized, Tanner2022Compressed}. Finally, in line 8, it estimates the current foreground element $\HTs{P^j - \alpha_j R^j}$ by projecting the gradient updated $P^j$ on the sparse positive constraint \RevT{using $\HTs{\cdot}$ which keeps only $s$ largest positive entries while setting others to zero.}}


To provide the algorithm with an initial guess, we first project the given data matrix $X$ on the low-rank constraint as $L^0 = \HTr{X}$. We then fit the least squares problem with $\hat{P}^0 = \argmin_P \left\| L^0 + \Psi(P) - X\right\|_F^2$ with a least-squares solver for sparse systems and project the result on the sparse constraint as $P^0 = \HTs{\hat{P}^0}$.

\begin{algorithm}[t]
\caption{NAHT algorithm for \eqref{eq:optimization}}\label{algo:naht}
\hspace*{\algorithmicindent} \textbf{Input:} $X\in\bR^{T\times N^2}$ and parameters $r,s\in\mathbb{N}$ 
\begin{algorithmic}[1]
\State $L^0 = \HTr{X}$
\State $\hat{P}^0= \argmin_P \left\| L^0 + \Psi(P) - X\right\|_F^2$
\State $P^0 = \HTs{\hat{P}^0},\, \Omega^0 = \mathrm{supp}(P^0),\,j = 0$
\While{not converged}
\State $L^{j+1} =  \HTr{\Rev{X- \Psi(P^j)}}$
\State $R^j =  \Rev{-\Psi^*\left(X - L^{j+1} - \Psi(P^j)\right)}$
\State $\alpha_j = \left\| R^j_{\Omega^{j}} \right\|_F^2 \big/ \left\|\Psi \left(R^j_{\Omega^{j}} \right)\right\|_F^2$
\State $P^{j+1}=  \HTs{P^j - \alpha_j R^j},\, \Omega^{j+1} = \mathrm{supp}(P^{j+1})$
\State $j = j+1$
\EndWhile
\end{algorithmic}
\end{algorithm}

\vspace{-1em}
\section{Numerical experiments}\label{sec:numerics}

\begin{figure*}[t!]
     \centering
     \begin{subfigure}[b]{0.295\textwidth}
         \centering
         \includegraphics[width=\textwidth]{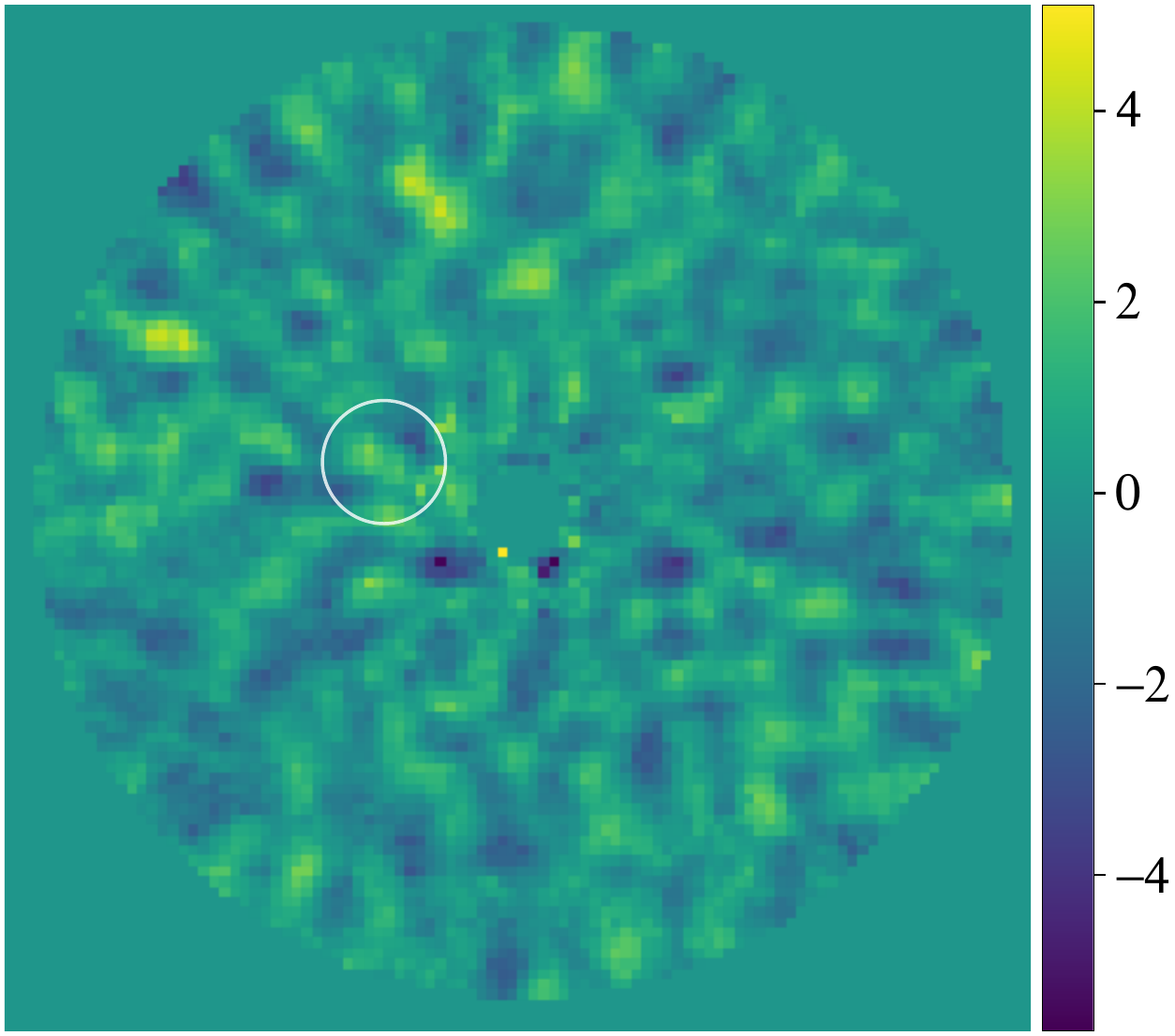}
         \caption{AnnPCA-SNR\label{fig:betapic_fake_snr}}
     \end{subfigure}
     \hfill
     \begin{subfigure}[b]{0.29\textwidth}
         \centering
         \includegraphics[width=\textwidth]{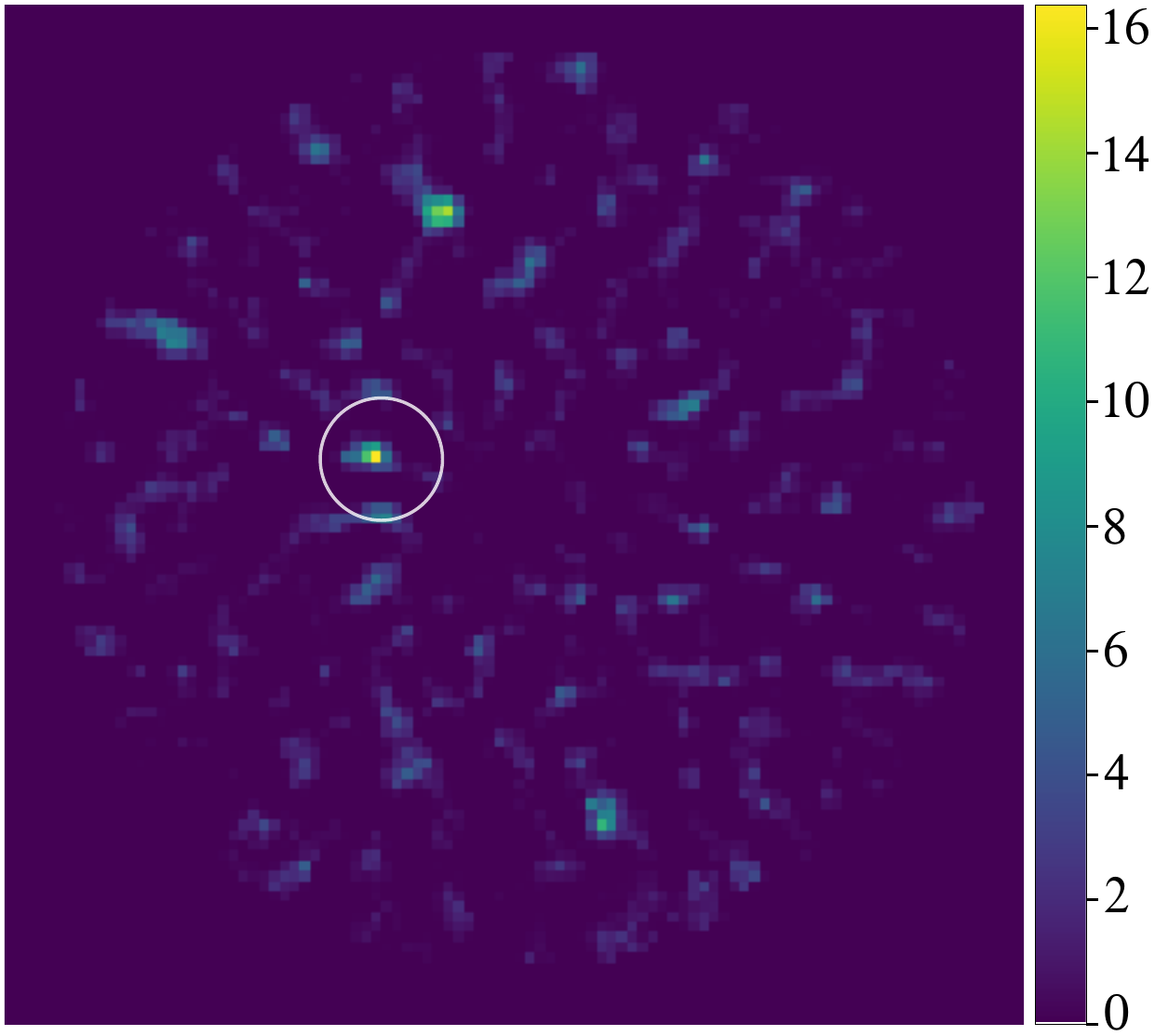}
         \caption{AnnPCA-LR\label{fig:betapic_fake_lr}}
     \end{subfigure}
     \hfill
     \begin{subfigure}[b]{0.3\textwidth}
         \centering
         \includegraphics[width=\textwidth]{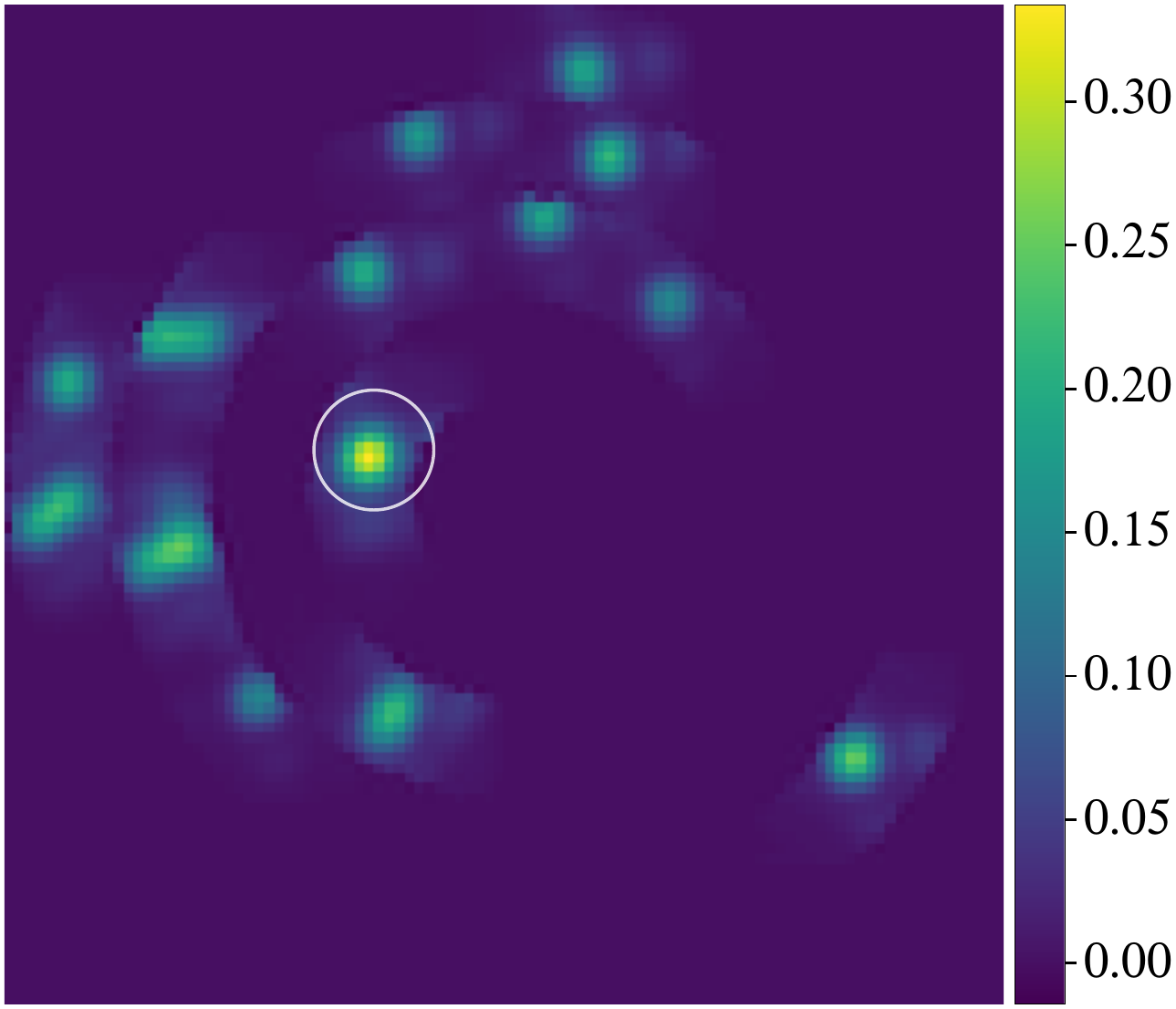}
         \caption{LRPT\label{fig:betapic_fake_lrpt}}
     \end{subfigure}
        \caption{Plot of the detection map for each of the methods from a single experiment on the $\beta$-Pictoris dataset with an injected planet at $3  \lambda/D$ seperation for flux $1.5\sigma_{ann}$, with the correct location circled in white. \Rev{The different color scales do not necessarily affect the ROC curves, since the detection threshold is chosen relative to the highest observed value.} \label{fig:betapic_detection}}
\end{figure*}

We provide an experimental comparison of the proposed \Rev{LRPT} method with two variants of the \Rev{widely used AnnPCA method}, one based on the SNR detection map (AnnPCA-SNR) \cite{GomezGonzalez2017Vip} and the other one on likelihood ratio (AnnPCA-LR) \cite{Daglayan2022Likelihood}. \Rev{We compare the methods} in terms of the visual quality of the detection maps and the Receiver operating characteristic (ROC) curves. The reproducible Python implementation is publicly available\footnotemark[1].

We perform the tests on the common benchmark of the ADI cube VLT/NACO $\beta$-Pictoris in the infra-red L' band (3.8{$\mu$m}), with 612 frames covering 83$^{\circ}$ of parallactic angles and $\lambda/D \approx 4.6\,\text{pixel}$ \cite{Absil2013Searching}. In order to reduce the computation time, we subsample the cube to include only every third frame, resulting in a cube of size $204 \times 100 \times 100$. 

To compare the methods, we set the rank in the methods to be the best performing rank in case of the flux of 1.5$\sigma_{ann}$, relative to the standard deviation of the pixel intensities in the annulus $\sigma_{ann}$, which \Rev{is equal to} $r=10$ for LRPT and \Rev{to} $r=35$ for AnnPCA. The sparsity parameter for LRPT is chosen to be $s=10$. In order to provide a fair comparison, we implement an annular version of the LRPT method by solving the optimisation problem \eqref{eq:optimization} for different annular regions as is also done in AnnPCA. Moreover, in AnnPCA we disable the additional heuristic for parallactic angle threshold in the VIP package \cite{GomezGonzalez2017Vip}. Turning off the angle heuristic ensures that the low-rank component is based on the same principle in both of the methods which results in the same model for the quasi-static field\footnote{This is achieved by setting \texttt{delta\_rot} parameter to zero and \texttt{max\_frames\_lib} parameter to the number of frames in the \texttt{vip\_hci.psfsub.pca\_local} module.}. 

Fig.~\ref{fig:betapic_detection} shows the visual quality of detection maps for the methods on the $\beta$-Pictoris dataset for a synthetically injected planet with the value of flux $1.5 \sigma_{ann}$ relative to the standard deviation of the pixel intensities in the annulus $\sigma_{ann}$. We see that whereas the AnnPCA-SNR fails to produce a detection map clearly identifying the exoplanet, AnnPCA-LR and LRPT manage to produce meaningful guesses. Moreover, LRPT produces a detection map based on a positive sparsity criterion in \eqref{eq:model} that could be further fine-tuned to detect a specific number of planets.

\begin{figure}[t !]
    \centering
    \includegraphics[width = 0.42\textwidth]{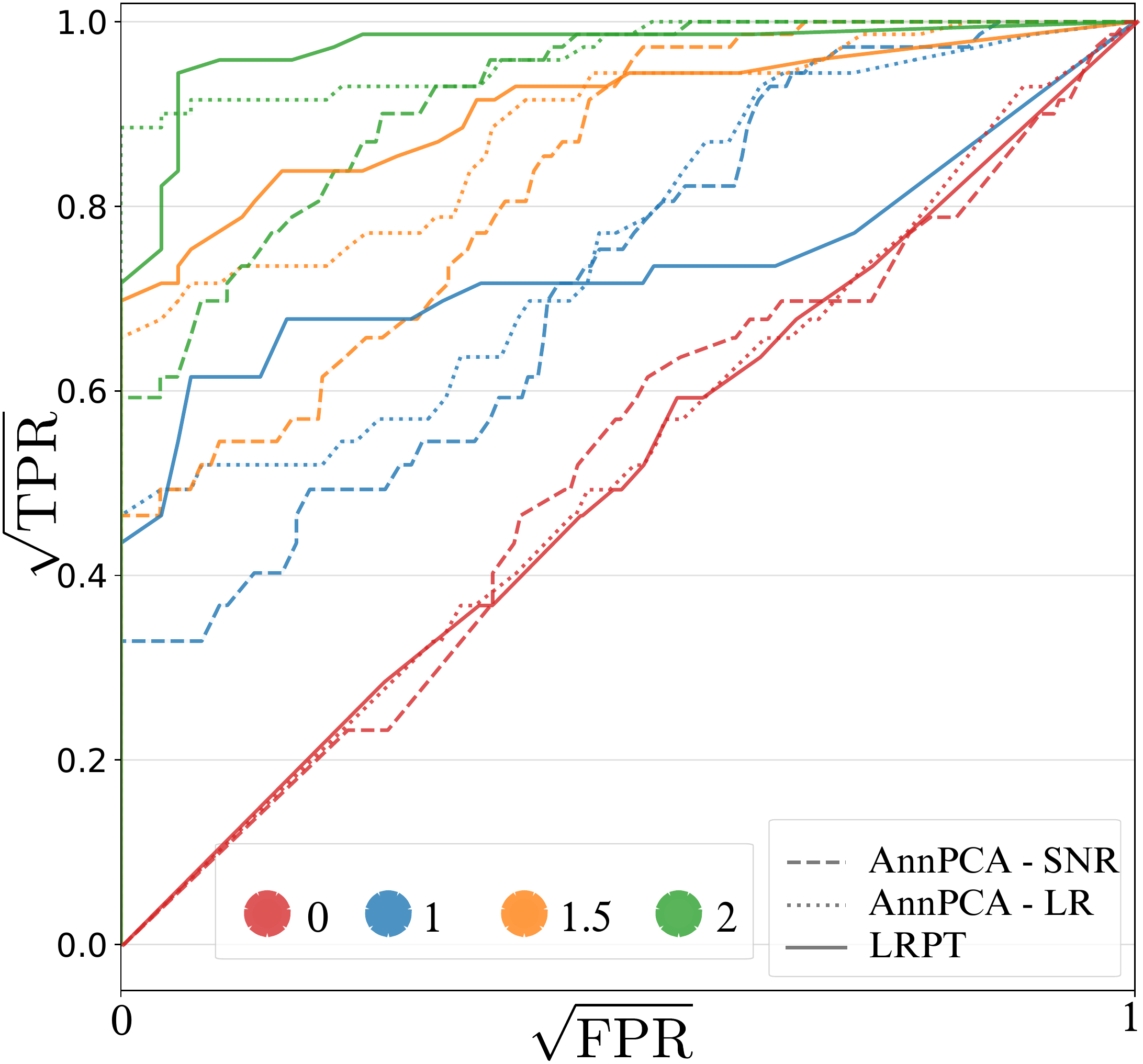}
    \caption{ROC curves for $\beta$ Pictoris data set. The planets are injected in 3 $\lambda/D$ separation with the fluxes indicated in the legend multiplied by the standard deviation of the annulus. In order to better observe the low FPR regime, we scale the axes using the square root.
    \label{fig:betapic_roc}}
\end{figure}

We compute the deterministic \RevT{receiver operating characteristic (ROC) curves using the methodology for counting the true positive (TPR) and false positive (FPR) detections} as in \cite{Daglayan2022Likelihood}, whose main advantage is that it is easily reproducible and that for datasets without a planet \Rev{any detection algorithm would} generate a diagonal line. To compute ROC curves, we generate synthetic groundtruth examples with varying flux, i.e., intensity of the planet, by injecting the planet-free data cube (the data cube where the known planets have been removed) with fake planets using the VIP HCI package \cite{GomezGonzalez2017Vip}. \RevT{Although the injection of different levels of quasi-static noise is not possible using the VIP HCI package, we change the signal to noise ratio by controlling the intensity of the injected planets.} Fig.~\ref{fig:betapic_roc} depicts the performance of the methods in terms of deterministic ROC curves computed for values of flux relative to the standard deviation of the annulus $c \cdot \sigma_{ann}$, where $c = 0, 1, 1.5, 2$. \Rev{In order to lower the computational time we crop the $\beta$-Pictoris dataset around the star resulting in a cube of size $204 \times 60 \times 60$.} We observe that the LRPT method consistently outperforms the AnnPCA-SNR for all tested values of flux and AnnPCA-LR for flux $c=1.5$. 

\section{Conclusion \& future work}\label{sec:conclusion}

We proposed a model-based method for exoplanet detection that attempts to capture both the structure of the quasi-static speckled field and the shape of the circular planet trajectories can take in concurrence. This opens several avenues for further research, for example, by imposing sparsity in different domains to detect other celestial objects, such as circumstellar discs \Rev{\cite{Pairet2021MAYONNAISEa,Soummer2012Detection}}. 
Finally, one could adapt this method to work on images with multitude of spectral bands \Rev{for which other geo-metric transformations, such as radial dilation, are expected}. 


\bibliographystyle{IEEEbib}
\bibliography{refs}

\end{document}